# IEEE Copyright Notice

# Transient Stability Assessment of Cascade Tripping of Renewable Sources Using SOS

Chetan Mishra, James S. Thorp, Virgilio A. Centeno
Bradley Dept. of Electrical and Computer Engineering,
Virginia Polytechnic Institute and State University,
Blacksburg-24061, Virginia, USA
Email: {chetan31, jsthorp, virgilio}@vt.edu

Anamitra Pal
School of Electrical, Computer, and Energy Engineering
Arizona State University
Tempe-85287, Arizona, USA
Email: Anamitra.Pal@asu.edu

*Abstract—* There has been significant increase in penetration of renewable generation (RG) sources all over the world. Localized concentration of many such generators could initiate a cascade tripping sequence that might threaten the stability of the entire system. Understanding the impact of cascade tripping process would help the system planner identify trip sequences that must be blocked in order to increase stability. In this work, we attempt to understand the consequences of cascade tripping mechanism through a Lyapunov approach. A conservative definition for the stability region (SR) along with its estimation for a given cascading sequence using sum of squares (SOS) programming is proposed. Finally, a simple probabilistic definition of the SR is used to visualize the risk of instability and understand the impact of blocking trip sequences. A 3-machine system with significant RG penetration is used to demonstrate the idea.

## I. INTRODUCTION

Renewable generation (RG) has introduced a high degree of complexity to power system operations. Besides the uncertainty associated with their intermittency, another phenomenon that has put RG-rich grids under severe risk of instability is the offline tripping of RGs during disturbances. Traditionally, RG sources were designed to *not* support the grid during disturbances and were made to trip offline. However, increasing numbers of RG sources has made utilities and inverter manufacturers realize that this will no longer be a viable design. This has led to the formulation of ride through curves [1] which define limits on frequency and voltage of RGs for staying online. The overall objective being to maintain high reliability by preventing RG-induced instabilities [2].

Isolated tripping of a few RGs will not pose a serious threat to system stability. However, the development of RG plants is primarily driven by geographical and economical factors, which result in some degree of electrical coherency among the plants. The problem occurs when localized tripping of coherent RG sources triggers a cascade. Furthermore, utilities are driven towards standardizing the ride through settings across their territory. This directly maps the electrical coherency to a tripping sequence – *a perfect setting for potential cascading failures*. As these cascading events happen in a matter of seconds, they do not give any time to the system operators to react. This highlights the need for detailed studies of the impacts of fast-paced cascade tripping of RG sources on transient stability of the system.

Considering the many possible variations in time and order of tripping, there can be infinitely large number of cascading scenarios. This prevents time domain simulations from becoming a practical approach for analyzing stability of RG-rich systems. The only alternative is the use of *direct methods*. Direct methods have significantly evolved in the last 30 years [3]. However, they were developed to deal with continuous systems and relied on the proper characterization of the stability boundary. A system prone to RG tripping has continuous dynamics but with discrete changes in it referred to as a *switched system*, which is fundamentally different. The stability of switched systems is defined with respect to a set of stable equilibrium points (SEPs) rather than a single one. In addition, characterizing the stability boundary for such systems is a very challenging task. Therefore, not all direct methods are equally successful in analyzing stability of switched systems. Lyapunov's direct method [4] provides a way of estimating the stability region (SR) using level set(s) of a scalar function satisfying certain criteria. This has been proven to be effective in dealing with stability of more complex dynamical systems including switched systems [5]. However, the main challenge with this technique is finding a suitable scalar function. With the advancements made in sum of squares (SOS) techniques, it has become possible to find such scalar functions using convex optimization [4]. Milano et al. [6] demonstrated the application of SOS programming for estimating the SR for a power systems classical model. Mishra et al. demonstrated the application of SOS programming for estimating SR under ride through constraints [7]. To the best of the authors' knowledge, no prior attempts have been made at understanding the implication(s) of fast-paced cascade tripping of RGs on system stability. The aim of this paper is to develop a framework for assessing the system stability under this phenomenon using Lyapunov's direct method and SOS programming.

The rest of the paper is structured as follows. In Section II, the concept of stability under a given cascading event scenario is presented. A conservative definition of SR for a fixed cascade sequence independent of switching time is also provided. The estimation is done using multiple Lyapunov level sets. The idea of probabilistic SRs is presented for combining SR of every possible cascading sequence. Section III solves for those Lyapunov functions for a power systems classical model with trip-able RGs (modeled as negative loads) using SOS programming. The methodology is applied to a 3-machine system in Section IV. The conclusions and future scope of work are identified in Section V.

## II. Stability During Cascading Events

### A. Switched System Representation of Power System

A system prone to discrete changes can be modeled as a switched system [8] of the form:

$$\dot{x} = f_\sigma(x) \quad (1)$$

In (1), $\sigma$ is a piecewise constant switching signal. Since we only consider tripping of RG sources as switching events, for a network with $n_{RG}$ number of RGs, we can have a maximum of $2^{n_{RG}}$ possible switching states. During a switching operation, the system's vector field changes along with its set of equilibrium points. Thus, *the stability is defined as settling to the relevant SEP of the final active vector field after all the switching has happened which is unknown beforehand*. Thus, the stability of the overall switched system is determined by the switching signal. Here it must be mentioned that according to the IEEE standard [1], once an RG has tripped, it cannot come back online until several minutes of normal operating conditions are established. This heavily limits the number of possible switching signals possible.

### B. Stability Region (SR) Under Multiple Switching Sequences

Transient stability assessment of RG-rich systems whose RGs are prone to tripping is a challenging problem. This is because the tripping of RGs is governed by multiple time dependent ride-through curves defined for each RG, which makes the switching signal both state and time dependent. The SR for such systems can be defined as a set of points in state-space from which the emerging trajectory converges to the SEP corresponding to the final switching state after all the switching has occurred. Unlike the case where stability is defined for a single system and it's SEP, this set is not necessarily *positive invariant*. This is because the final desired settling point (SEP of final switching state) for each cascading sequence can be different. Hence, it may not be possible to define a set of points in which the trajectories stay confined. Writing the overall conditions satisfied by this region in state-space is also not practical, as it will require formulating the constraints at the time and state at which the switching occurs. An added concern for system planners is that the cases with false tripping are also common, implying that there is a need to understand the consequences of certain unforeseen tripping events along each trajectory. Therefore, we will try to address the broader problem of estimating the stability under all likely switching sequences.

The stability along a trajectory is not only governed by the order of tripping of RGs but also the timing of the trips. This results in infinitely many switching signals to study for each trajectory making the problem impractical to deal with. Additionally, having a time element does not provide any useful information to the system planner when trying to understand the consequences. Therefore, in this paper, we study the worst-case scenario in terms of time elapsed between switching for each switching sequence. This results in the number of scenarios becoming finite, and the problem computationally tractable. For example, for $n_{RG}$ number of RGs, the number of possible switching signals would be $\sum_{r=0}^{n_{RG}} n_{RG}C_r \times r!$ where each switching signal is now only defined by which RGs trip and in what order in time.

### C. Stability Region (SR) for Fixed Cascading Sequence

Now, the first objective is to find the previously defined SR for a given fixed switching sequence, i.e. a region in state-space the trajectories starting from which converge to the SEP of the final switching state. In this paper, a switching signal with $N-1$ switchings will be denoted by a vector $\sigma = [\sigma_1, \sigma_2 ... \sigma_N]$ where $\sigma_i$ gives the switching state (connection status of all the RGs) before the $i^{th}$ switching takes place for this cascade/ switching signal. Please note that for the rest of the paper, the terms switching signal and cascading sequence have been used interchangeably. We now propose the following Theorem to estimate the desired SR.

**Theorem 1:** Let $\Omega_j$ denote a positively invariant set for the system $\dot{x} = f_j(x)$ s.t. $\Omega_j \subseteq A_j(x_{sep}^{(j)})$ where $x_{sep}^{(j)}$ is the relevant SEP and $A_j(x_{sep}^{(j)})$ is its SR. Similarly, define $\Omega_k$ for the system $\dot{x} = f_k(x)$. If $\Omega_j \subseteq \Omega_k$, then the system is stable if it starts in $\Omega_j$ with status $j$ active and any of the following scenarios happen: (1) no switching; (2) switching from $j \to k$.

**Proof:** For the no switching case, it holds true by definition. For the switching case, the trajectory starting in $\Omega_j$ will stay within it till the active vector field is $f_j(x)$ due to its positively invariant nature. Since $\Omega_j \subseteq \Omega_k$, the trajectory is also bounded inside $\Omega_k$. Now, as soon as switching to $k$ happens, the vector field $f_k(x)$ activates and by definition $\Omega_k$ is a positively invariant subset of its SR. Thus, the trajectory eventually converges to $x_{sep}^{(k)}$.

For a given $\sigma$, the SR estimate will be defined by the innermost set $\Omega_{\sigma_1}$. In our work, we propose estimating each set $\Omega_{\sigma_i}$ by a corresponding Lyapunov function's level set $\{V_{\sigma_i} \leq 1\}$ which needs to satisfy the following conditions:

$$V_{\sigma_i}(x) > 0, V_{\sigma_i}(0) = 0$$
$$\frac{\partial V_{\sigma_i}(x)}{\partial x} f_{\sigma_i}(x) < 0 \; \forall x \in \{x \neq 0 | V_{\sigma_i}(x) \leq 1\} \quad (2)$$
$$V_{\sigma_{i+1}}(x) \leq 1 \; \forall x \in \{x | V_{\sigma_i}(x) \leq 1\}$$

### D. Probabilistically Combining Stabilty Region (SR) Estimates for Cascading Sequences

We propose the concept of *risk of instability region* as an indicator of the likelihood of instability under plausible cascade tripping sequences. Since utilities keep a record of various RG tripping events, this record can be used to estimate the relative frequency of each cascading sequence. If $p_i$ be the probability of $i^{th}$ cascading sequence $\sigma^i$, then the risk of instability for a trajectory starting at $x_0$ can be written as:

$$Risk(unstable|x_0) = \sum_i p(unstable|x_0|\sigma^i) \times p_i \quad (3)$$

In (3), the first term being multiplied on the RHS will be 1 if $x_0$ is outside the estimated SR for $\sigma^i$, and 0 otherwise, while $p_i$ can be chosen according to the preference of the operator. The concept of risk of instability region helps gain an insight into the impact of RG tripping and can identify key events that can put the system under risk. These events can then be prevented by choosing an optimal blocking logic. For example, consider a system with three RGs for which the following blocking logic has been defined: RG 1 can trip only if RG 2 has not tripped, and vice-versa. Then, on implementing this blocking logic, if the original cascading sequence was 1-2-3, the modified sequence will be 1-3. The following assumptions were made while implementing this logic:

1. If $\sigma^i$ transforms to $\sigma^j$, then:
    a. $p^{new}(\sigma^j) = p^{old}(\sigma^j) + p^{old}(\sigma^i)$
    b. $p^{new}(\sigma^i) = 0$
2. Blocking the tripping of an RG during a cascade does not change the way the rest of the cascade propagates, i.e. each $\sigma^i$ has a single $\sigma^j$ that is known beforehand.

## III. APPLICATION OF SUM OF SQUARES (SOS) TO STABILITY REGION (SR) ESTIMATION FOR CLASSICAL MODEL

### A. Power Systems Classical Model under RG Tripping

In this study, a classical model for the power system is used with the RGs modeled as negative real loads. On reducing the network buses, the final network impedances will be a function of the connection status of all the RGs (switching state). Thus, for a given switching signal $\sigma$, the power system classical model in one machine reference ($n_g^{th}$ generator) frame with uniform damping can be written as,

$$\dot{\delta}_{in_g} = \omega_{in_g}$$

$$M_i \dot{\omega}_{in_g} = P_{m_i} - \left( \sum_{j=1:n_g} E_i E_j \left( G_{ij}(\sigma) \cos(\delta_{in_g} - \delta_{jn_g}) + B_{ij}(\sigma) \sin(\delta_{in_g} - \delta_{jn_g}) \right) + -\frac{M_i}{M_{n_g}} (P_{m_{n_g}} - \sum_{j=1:n_g} E_{n_g} E_j \left( G_{n_gj}(\sigma) \cos(-\delta_{jn_g}) + B_{n_gj}(\sigma) \sin(-\delta_{jn_g}) \right) \right) - D_i \omega_{in_g} \quad (4)$$

$$\delta_{in_g} = \delta_i - \delta_{n_g}, \omega_{in_g} = \omega_i - \omega_{n_g}, i = 1,2 \ldots (n_g - 1)$$

Since we will be using a sum of squares (SOS) based approach that requires a polynomial system, the first task is to convert the aforementioned system into a polynomial system through variable transformation. This is done as shown in the table below [6].

**Table 1 Power system variable transformation**

| New Variable | Function of Original States |
| --- | --- |
| $z_i^{(\sigma)}$ | $\omega_{in_g} - \omega_{in_{g_{sep}}}^{(\sigma)}(=0) = \omega_{in_g}$ |
| $z_{n_g+2i-2}^{(\sigma)}$ | $\sin(\delta_{in_g} - \delta_{in_{g_{sep}}}^{(\sigma)})$ |
| $z_{n_g+2i-1}^{(\sigma)}$ | $1 - \cos(\delta_{in_g} - \delta_{in_{g_{sep}}}^{(\sigma)})$ |

The superscript $(\sigma)$ used for the transformed states $z$ indicates that the variable transformation used is different for different switching states. The transformed polynomial system is given by,

$$\dot{z}^{(\sigma)} = f_\sigma(z^{(\sigma)}) \quad (5)$$
$$0 = g_\sigma(z^{(\sigma)})$$

Mapping between the polynomial states $z$ for different switching states $\sigma_j$ and $\sigma_k$ is as follows:

$$z_i^{(\sigma_j)} = z_i^{(\sigma_k)}$$

$$\begin{bmatrix} z_{n_g+2i-2}^{(\sigma_j)} \\ z_{n_g+2i-1}^{(\sigma_j)} \end{bmatrix} = [M_j] \times [M_k]^{-1} \times \left( \begin{bmatrix} z_{n_g+2i-2}^{(\sigma_k)} \\ z_{n_g+2i-1}^{(\sigma_k)} \end{bmatrix} - \begin{bmatrix} 0 \\ 1 \end{bmatrix} \right) + \begin{bmatrix} 0 \\ 1 \end{bmatrix} \quad (6)$$

$$[M_j] = \begin{bmatrix} \cos(\delta_{in_{g_{sep}}}^{(\sigma_j)}) & -\sin(\delta_{in_{g_{sep}}}^{(\sigma_j)}) \\ -\sin(\delta_{in_{g_{sep}}}^{(\sigma_j)}) & -\cos(\delta_{in_{g_{sep}}}^{(\sigma_j)}) \end{bmatrix}$$

### B. SR Estimate for a Fixed Cascade Sequence using SOS

SOS serves as a powerful tool to solve the problem described in (2). In this study, for each given switching signal $\sigma$ having the following transition of nested regions $\Omega_{\sigma_1} \to \Omega_{\sigma_2} \ldots \to \Omega_{\sigma_N}$, we sequentially estimate $\Omega_{\sigma_i}$ starting from the one corresponding to the last switching status ($\Omega_{\sigma_N}$) and working backwards. The methodology proposed in [6] is used to estimate each $\Omega_{\sigma_i}$ which is represented by $\{z^{(\sigma_i)} | V_{\sigma_i}(z^{(\sigma_i)}) \le 1\}$. Here it should be kept in mind that the function $V_{\sigma_i}$ would be estimated in terms of its own polynomial states $z^{(\sigma_i)}$. This means that when constraining $\Omega_{\sigma_i}$ to be inside $\Omega_{\sigma_{i+1}} = \{z^{(\sigma_{i+1})} | V_{\sigma_{i+1}}(z^{(\sigma_{i+1})}) \le 1\}, V_{\sigma_{i+1}}(z^{(\sigma_{i+1})})$ needs to be represented in terms of $z^{(\sigma_i)}$ (written as $V_{\sigma_{i+1}}(z^{(\sigma_i)})$) which can be done using (6). The initial estimate of $V_{\sigma_i}(z^{(\sigma_i)})$ is found by solving the SOS optimization problem shown in (7). Here we have assumed that all the functions involved in the SOS problems are functions of $z^{(\sigma_i)}$ and therefore have not written them explicitly. Also, $s_i$ would represent multiplier functions that are SOS themselves.

$$\max_{V_{\sigma_i}, \dot{V}_{\sigma_i}, s_2, s_6, \lambda_1, \lambda_2} \beta$$
$$-s_2(\beta - p) + V_{\sigma_i} - \lambda_1^T g_{\sigma_i} - l_1 \text{ is SOS} \quad (7)$$
$$-s_6(\beta - p) - \dot{V}_{\sigma_i} - \lambda_2^T g_{\sigma_i} - l_2 \text{ is SOS}$$

In (7), $\dot{V}_{\sigma_i} = \frac{\partial V_{\sigma_i}}{\partial z^{(\sigma_i)}} f_{\sigma_i}(z^{(\sigma_i)})$. Next, a local estimate of $\Omega_{\sigma_i}$ is obtained by solving the SOS optimization problem shown in (8), which forces $\Omega_{\sigma_i}$ to be inside $\{z^{(\sigma_i)} | V_{\sigma_{i+1}}(z^{(\sigma_i)}) \le 1\}$,

$$\max_{s_1, s_2, s_3, s_4, \lambda_1, \lambda_2} c$$
$$-s_1(-c + V_{\sigma_i}) - s_2(p - \beta_1) + s_3(-c + V_{\sigma_i})(p - \beta) - \lambda_1 g_{\sigma_i} - (p - \beta)^2 \text{ is SOS}$$
$$-s_4(c - V_{\sigma_i}) - \lambda_2 g_{\sigma_i} - (V_{\sigma_{i+1}} - 1) \text{ is SOS} \quad (8)$$
$$V_{\sigma_i} = \frac{V_{\sigma_i}}{c}, \dot{V}_{\sigma_i} = \frac{\dot{V}_{\sigma_i}}{c}$$

The final step is expanding this local estimate of $\Omega_{\sigma_i}$ using expanding interior algorithm [4] where a semi-algebraic set $P_\beta = \{p(z^{(\sigma_i)}) \le \beta\}$ for positive semi-definite function $p$ is expanded while constrained to be contained inside the estimate $\{V_{\sigma_i} \le 1\}$ with $V_{\sigma_i}$ being allowed to change. The standard formulation [6] when estimating the SR for a power system classical model is as follows,

$$\max_{s_2, s_6, s_8, s_9, \lambda_1, \lambda_2, \lambda_3} \beta \quad (9)$$
$$s_2 V_{\sigma_i} - \lambda_1^T g_{\sigma_i} - l_1 \text{ is SOS}$$
$$-s_6(\beta - p) - \lambda_2^T g_{\sigma_i} - (V_{\sigma_i} - 1) \text{ is SOS}$$
$$-s_8(1 - V_{\sigma_i}) - s_9 \dot{V}_{\sigma_i} - \lambda_3^T g_{\sigma_i} \text{ is SOS}$$

However, in our case, this estimate is also constrained to be inside $\{z^{(\sigma_i)} | V_{\sigma_{i+1}}(z^{(\sigma_i)}) \le 1\}$. This additional constraint needs to be written in the form of set emptiness condition from which we will derive the equivalent SOS conditions as shown below.

Inside $\Omega_{\sigma_i}$ estimate, $V_{\sigma_{i+1}}$ cannot be greater than 1: $\{V_{\sigma_i} \leq 1, V_{\sigma_{i+1}} \geq 1, V_{\sigma_{i+1}} \neq 1, g_{\sigma_i} = 0\} = \emptyset$. Using P-Satz Theorem [4], the constraint can be converted to: $s_{10} + s_{11}(1 - V_{\sigma_i}) + s_{12}(V_{\sigma_{i+1}} - 1) + s_{13}(V_{\sigma_{i+1}} - 1)(1 - V_{\sigma_i}) + (V_{\sigma_{i+1}} - 1)^{2k_1} + \lambda_4 g_{\sigma_i} = 0$. Finally, assuming $s_{10} = s_{11} = 0, k_1 = 1, \lambda_4 = (V_{\sigma_{i+1}} - 1) \times \lambda_4$, we get,

$$-s_{13}(1 - V_{\sigma_i}) - \lambda_4 g_{\sigma_i} - (V_{\sigma_{i+1}} - 1) = s_{12} \text{ is SOS} \quad (10)$$

The final estimate heavily depends on the choice of function $p$ which is overcome up to a certain extent by the methodology proposed in [6]. The idea is that for a fixed $p$, $P_\beta$ is expanded and once it converges, $p$ is replaced by the obtained value of $V_{\sigma_i}$, and the optimization is re-run. This step is repeated until $P_\beta$ converges to $\{V_{\sigma_i} \leq 1\}$.

## IV. RESULTS

The proposed methodology is demonstrated on a 3-machine system that contains three RGs which amounts to an overall penetration of ~50%. Machine 3 is the reference and the system is represented in single machine reference frame. A uniform damping $\frac{D}{M}$ value of 4 is used for the simulations.

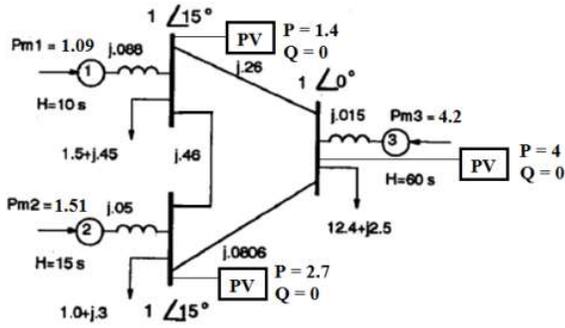

**Fig. 1: 3-Machine test case**

We use quadratic Lyapunov functions to estimate the SRs since higher degree options are not tractable. The following degrees are chosen for the various multiplier functions in the SOS optimization: $d(s_2) = 0, d(s_6) = 0, d(s_8) = 2, d(s_9) = 0, d(s_{13}) = 0, d(\lambda_1) = 0, d(\lambda_2) = 0, d(\lambda_3) = 0, d(\lambda_4) = 0$. For more details on choice of degrees, please refer to [4]. There are $2^3 = 8$ switching states. The corresponding SEPs are given in Table 2 with possible switching signals in Table 3.

**Table 2: SEP for various switching states**

| Switching State | RG Connection Status (1-online, 0-offline) | SEP $(\delta_{13}, \delta_{23})$ | Switching State | RG Connection Status (1-online, 0-offline) | SEP $(\delta_{13}, \delta_{23})$ |
|---|---|---|---|---|---|
| 1 | 111 | (0.2916, 0.2741) | 5 | 011 | (0.0767, 0.2567) |
| 2 | 110 | (0.4000, 0.3362) | 6 | 010 | (0.1790, 0.3177) |
| 3 | 101 | (0.2885, 0.1124) | 7 | 001 | (0.0731, 0.0950) |
| 4 | 100 | (0.3976, 0.1714) | 8 | 000 | (0.1754, 0.1532) |

**Table 3: Possible switching signals**

| Switching Signal | RG Trip Order | Switching State Order (refer to Table 2) | Switching Signal | RG Trip Order | Switching State Order (refer to Table 2) |
|---|---|---|---|---|---|
| 1 | No trip | 1 | 9 | 3 2 | 1 2 4 |
| 2 | 1 | 1 5 | 10 | 2 3 | 1 3 4 |
| 3 | 2 | 1 3 | 11 | 3 2 1 | 1 2 4 8 |
| 4 | 3 | 1 2 | 12 | 3 1 2 | 1 2 6 8 |
| 5 | 2 1 | 1 3 7 | 13 | 2 3 1 | 1 3 4 8 |
| 6 | 1 2 | 1 5 7 | 14 | 2 1 3 | 1 3 7 8 |
| 7 | 3 1 | 1 2 6 | 15 | 1 2 3 | 1 5 7 8 |
| 8 | 1 3 | 1 5 6 | 16 | 1 3 2 | 1 5 6 8 |

Let us take the example of estimating SR for switching signal 16 using the methodology proposed in Section III.B. For this case, $\sigma$ takes the following switching states: 1-5-6-8. Fig. 2 gives the nested sets with the innermost blue set depicting the SR estimate. It is important to mention here that there can be cascade sequences where it is not possible to find such uniformly nested regions. Such sequences have high risk of instability, and are precisely the sequences that a planner would want to avoid by implementing a clever blocking logic.

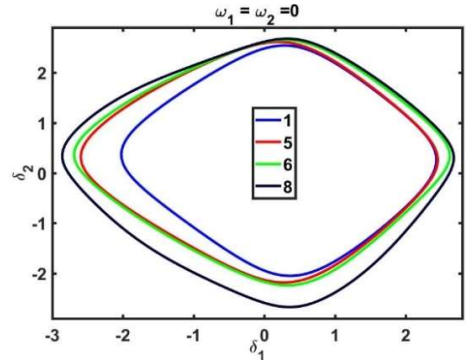

**Fig. 2: Nested invariant sets for switching signal 16**

Next, for time domain validation, the trajectories are initiated at the boundary of the blue region (switching state 1) and switched at random intervals sampled from $[0, 2]$ seconds in the order $1 \rightarrow 5 \rightarrow 6 \rightarrow 8$. The projection of those trajectories on the angle plane is shown in Fig. 3. The figure confirms that the trajectories converge to the SEP corresponding to the final switching state (which in this case is 8). Each trajectory has four colors that represent the active vector field (see Fig. 2 for color code). Similarly, plotting the SR estimate for all 16 switching signals, we get Fig. 4. Let us now combine these estimates probabilistically by assuming that all these sequences are equally likely (i.e. $p_i = \frac{1}{16}$). The resulting risk of instability is plotted in Fig. 5. The risk of instability for different starting points in state space increases from blue to red with blue being the lowest. It can be seen that there is a significantly sized region (colored blue) which has 0 risk. This region could be attributed to the largely overlapping SR estimates for each sequence.

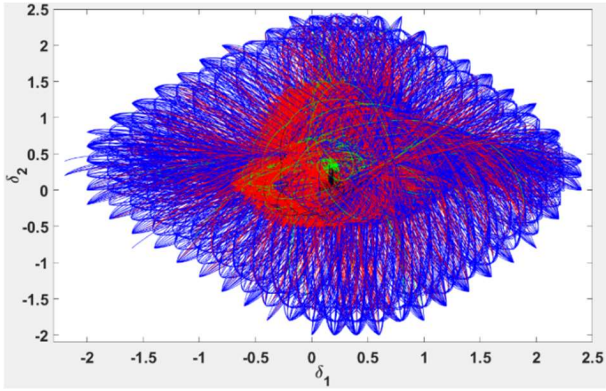

**Fig. 3: Trajectories starting inside SR of switching signal 16**

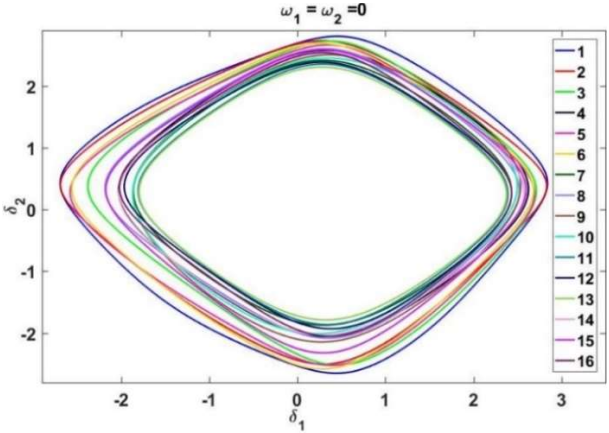

**Fig. 4: SR estimates for all switching signals**

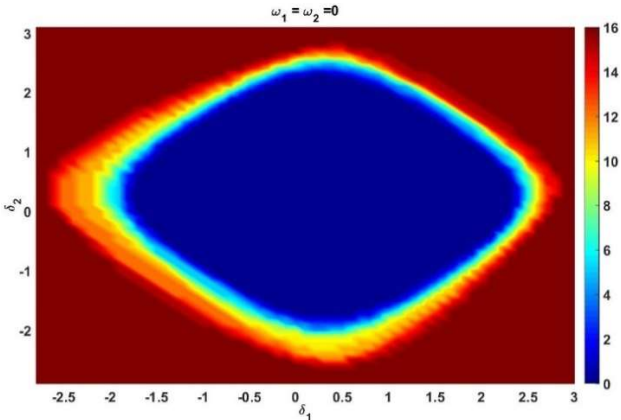

**Fig. 5: Risk of instability region**

Let us now visualize the impact of blocking some tripping sequences on the risk of instability regions. Here, we consider a blocking logic where the selected RGs check each other's connection statuses such that when the blocking logic is implemented, the last one remaining online does not trip. For example, if blocking logic $B = [1, 2]$ is implemented, then RG 1 will not trip if RG 2 has tripped, and vice-versa. From Fig. 6 it can be observed that for the 3-machine test system, blocking the tripping of RG 3 has the best reduction in the risk of instability (indicated by the biggest blue region in the top right image) as compared to other options.

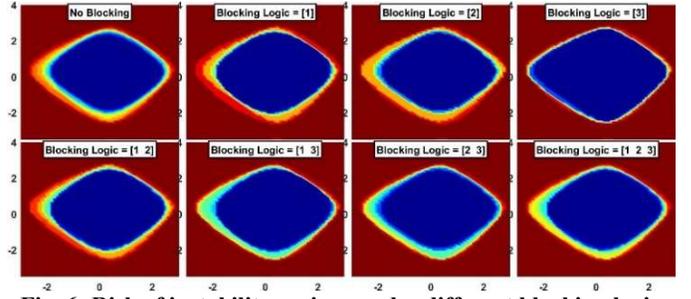

**Fig. 6: Risk of instability regions under different blocking logic**

## V. CONCLUSIONS AND FUTURE WORK

This paper proposes a direct method based methodology to understand the impact of cascade tripping of RGs on transient stability. The power system is modeled as a switched system. A definition of the SR that is robust against switching times for a fixed cascade sequence is proposed. The region is estimated using vector Lyapunov functions found using SOS programming. To understand the overall impact of multiple cascaded trips on stability, the concept of risk of instability is introduced, which highlights the consequences of blocking the tripping of certain RG sources during disturbances. The proposed approach is demonstrated on a multi-machine classical model with significant RG penetration.

This idea is still in its conceptual phase with many directions to be explored in the future. For instance, although for smaller systems, the SRs can be visualized, for larger systems, it is difficult to relate them to stability. Therefore, there is a need for physically relatable scalar metrics that can reflect the increase in risk due to multiple cascade sequences. Another area that can be further explored is the reduction in conservativeness of SR estimate for fixed cascades. Lastly, the computational challenges associated with applying SOS based techniques to large systems also needs to be solved.